\documentclass[aps,prl,preprint,groupedaddress,showpacs]{revtex4-1}
\usepackage{graphicx}
\usepackage{float}
\usepackage{epsfig}
\usepackage{natbib}

\begin{document}


\title{Topography induced optical spectral shifts and finite size effect of focal spot}


\author{V. Tishkova and W.S.Bacsa}
\email[]{wolfgang.bacsa@cemes.fr}
\affiliation{Centre d'Elaboration de Mat\'{e}riaux et d'Etudes Structurales du CNRS, Universit\'{e} de Toulouse, 29 Jeanne Marvig, 31055 Toulouse, France}


\date{\today}

\begin{abstract}
We observe topography induced spectral shifts using high resolution grating spectrometers which we attribute to the fact that the focal spot has a finite size. The topography induced spectral shifts depend on spectrometer grating orientation and numerical aperture of the microscope objective. This is demonstrated by spectroscopic imaging trenches in GaAs in directions parallel and perpendicular the spectrometer entrance slit. Differences along the two directions of the LO phonon band show that the spectral shift is due to the variation of the grating angle across the non uniform illuminated focal spot caused by topography. Alignment errors of the optical axis lead to additional spectral shifts. Topography induced spectral shifts can be detected by recording spectra by scanning the sample in two perpendicular orientations with respect to the spectrometer entrance slit.
\end{abstract}


\maketitle

\section{}
\paragraph{}
Surface topography and composition of semiconductors can be controlled at scales which are one order of magnitude smaller than the optical wavelength. In order to use the spectroscopic information, it is important to know the influence of external parameters on the recorded spectra. We show here that spectroscopic shifts can be observed from the non-uniform illumination of the focal spot due to topography. The size of the focal spot is defined by the point spread function and is limited by diffraction. This means in lateral direction the focal spot is limited by half the wavelength and along the optical axis by two times the wavelength \cite{abbe,born}. Far field optical microscopy takes advantage of ray optics while spectroscopy takes advantage of diffractive optics. Ray and diffraction optics are combined when using spectroscopy at microscopic scales. We show here that the two are intrinsically related at sub-wavelength scales. Spectroscopic signals depend in general not on topography. However, we find that at scales smaller than the size of the focal point, topography can influence the detected spectroscopic signal when using grating spectrometers. 
\paragraph{}
	Recent findings on emission of individual carbon nanotubes have shown that the spectral position of the detected light emission depends on the exact location of the nanotube within the focal spot and can be as large as 5 $cm^{-1}$ depending on numerical aperture, grating dispersion and orientation \cite{Walsh2008}. An individual nanotube with a diameter two orders smaller than the focal spot represents an ideal point source. This gives the opportunity to use individual nanotubes to probe the optical focal spot and to test optics for scales smaller than the focal spot. The spectroscopic shift as a function of location of the nanotube in the focal spot has been explained by the fact that off axis optical emission results in a change in the grating angle, the angle between the incident and on the grating diffracted beam. As a result a spectroscopic shift is measured when using high resolution optical spectrometers. We show here that this effect is not unique to individual nano emitters but is present whenever the focal spot is not illuminated uniformly such as for topography variations smaller than the size of the focal spot. We use topography variations in the form of well aligned trenches fabricated by electron beam lithography. The width of the edges is smaller then the focal spot size ($<100nm$) leading to a non uniform emission and illumination of the focal spot and resulting in measurable spectral shifts. Using fluorescence excitation spectroscopy it has been earlier shown that it is possible to localize single molecules well within the focal spot by spectrally selective imaging \cite{Oijen1998,Burns1985} by using the slightly different emission due to the molecular local environment. Several other concepts using far field fluorescence microscopy taking advantage of two photon excitation and time resolved readout have been recently shown to be able to break the diffraction limit \cite{Hell2004, Hell2007}. We show here that different locations within the focal spot can be discriminated using spectroscopy itself and without using fluorescence microscopy. We concentrate here on the relation between topography and detected spectral shift using high resolution grating spectrometers. 
\paragraph{}
     Spectral shifts as a function of different locations on substrates have so far been attributed to the effect of local strain \cite{Kisielowski1996}, phonon confinement \cite{Sui1992}, doping \cite{chen2002} and structural changes \cite{Jiao2009}. To clearly discriminate spectral shifts associated to these effects, we scan the substrate in two perpendicular directions with respect to the spectrometer entrance slit and compare the recorded spectra as a function of exact position on the sample. In the case of individual carbon nanotubes the observed spectral shifts have been shown to be proportional to the total optical magnification and to be linear with the dispersion of the spectrometer \cite{Walsh2008}. The  exact position of the focal point with respect to the substrate influence the way the optical signal is detected. The  optical path of the emitted light is also affected by the alignment of the optical microscope with respect to the spectrometer optical axis. This means that any alignment errors of the microscope with respect to the spectrometer results to additional spectral shifts. We demonstrate these two effects using trenches in a GaAs waver and record multiple spectra in a direction always perpendicular to the trenches and either parallel or perpendicular to the grating direction. Figure \ref{fig1} shows that the two scan directions with respect to the spectrometer grating. For both configurations the polarization is set parallel to the spectrometer grating direction. Figure \ref{fig1} shows also a scanning force microscope image of the sample topography. Figure \ref{fig2} shows the spectroscopic position of the LO phonon energy of GaAs (symmetry E) \cite{Parayanthal} when scanning the sample perpendicular to the direction of the trenches and orienting the sample either perpendicular or parallel to the spectrometer grating (top/bottom) and using 413nm laser excitation (spectrometer Horiba T6400, grating: 2400 lines/mm). The sample has been scanned in steps of 50 nm and we have used a numerical aperture of 0.9. The spectral peak position when scanning in a direction perpendicular to the spectrometer entrance slit shifts on average by 0.53 $cm^{-1}$ when scanning over the etch of the trenches. When scanning parallel to the spectrometer slit the topography induced shifts are consistently smaller and oscillate on average 0.21 $cm^{-1}$ or less. The spectroscopic position as shown in the figure \ref{fig2} has been determined by fitting each experimental spectra with a Lorentzian lineshape. The differnces are explained by the fact that topography leads to a non-uniform illuminated focal spot and the optical axis is as a result displaced by a fraction of the size of the focal spot.
\paragraph{} 
	When repeating the experiment we noticed that alignment errors of the microscope with respect to the optical axis of the spectrometer or tilt of the sample lead to additional spectral shifts. It is clear that any deviation from the optical axis implies a change of the grating angle and as a result the detected emission is shifted in proportion to the alignment error. This explains that there is also a small topography induced shift observed for the direction parallel to the spectrometer grating direction. However, the topography induced shift is clearly larger when scanning the substrate perpendicular to the entrance slit. The here observed spectroscopic shift is smaller than what has been observed for individual carbon nanotubes ($<5 cm^{-1}$) in a similar experimental situation \cite{Walsh2008}. This is explained by the fact that the sample surface is extended and larger that the focal spot size. At the edge one side of the focal spot is illuminated unlike for individual nanotubes where onlt a small fraction of the focal spot is illuminated. Furthermore large spectral shifts are associated with smaller spectral intensity contributing to a lesser degree to the total signal. 
\paragraph{}
	Figure \ref{fig2} shows the intensity of the LO phonon band when scanning perpendicular (left side) and parallel (right side) to the spectrometer entrance slit. The intensity variations are large when the spectrometer slit is perpendicular to the trenches and absent when parallel. The intensity varies by a factor 5 when scanning perpendicular to the spectrometer entrance slit and the maximum intensity is the same for both sample orientations. We find that the relative phase of the oscillations of peak position and peak intensity with displacement are influenced by the exact alignment.
\paragraph{}
	Topography changes imply that the surface is either higher or lower with respect to the focal spot. Figure \ref{fig3} shows the effect of over or under focusing of the objective lens on the recorded spectra. When the focal spot is too high, the oscillations of the spectral position is larger (0.95 $cm^{-1}$) compared to when the focal point is on the surface (sample in perpendicular orientation). When the focal spot is too low, the oscillations of the spectral position are barely visible. Clearly the exact spectral position is also influenced by the vertical position of the focal spot with respect to the surface. Focusing errors lead to an enlargement of the focal spot on the sample. We have selected a focal spot size when over or under focusing, which is double in size compared to when in focus. A larger focal spot and non-uniform emission leads to a larger change in the grating angle and resulting spectroscopic shifts. To prevent any spectroscopic shifts due to focusing it is important to orient the substrate precisely perpendicular to the optical axis. Slopes of the substrate surface lead to changes in focusing and this influences the detected spectroscopic position. In the case of alignment errors, the under focused or over focused spot is displaced with respect to the optical axis by a different amount and this can explain the differences in the observed spectral shifts. While alignment errors of the microscope and tilt of the surface with respect to the optical axis can in principle be corrected, the spectral shift due the finite size of the focal spot and topography is intrinsic. Topography induced shifts can be detected by recording spectra in the same area for two sample orientations in a plane perpendicular to the optical axis. Any differences in the spectral position indicate that the topography variation along the two directions is not the same. The frequency shifts observed in figure \ref{fig1} are qualitatively the same as in figure \ref{fig3} showing that the main effect is due to height variations due to topography. There are remaining questions about the phase shifts between variations in spectroscopic position and intensity as well as details about the variation at the edge of the trenches. We have here used  spectral shifts of a phonon band of the substrate. It is important to note that the elastic peak, the emission of the laser line, can be used in the same way.    
\paragraph{}
	We have demonstrated that topography can lead to spectroscopic shifts and strong intensity variations which is attributed to the fact that the focal spot is not illuminated uniformly. The exact location of the focal spot with respect to the sample surface, as well as alignment errors influence the spectral position. While alignment errors of the optical spectrometer, microscope and laser beam can be corrected, variations or non uniformity of the collected optical signal within the size of the focal spot due to topography are intrinsic. This effect can be reduced with under focusing at the expense of recording lower signals or measuring in a direction parallel to the spectrometer slit. Topography at scales smaller than the size of the focal spot can be detected when comparing spectroscopic maps of different sample orientations with respect to the spectrometer entrance slit. The spectral shifts are limited by the size of the focal spot, the numerical aperture and the dispersion of the grating spectrometer. This shows that diffractive optics as used in grating spectrometers can be applied to obtain information about topography at scales below the size of the focal spot.

\begin{acknowledgments}
	\subsection{Acknowledgments}
 We would like to thank Benjamin Dwir from Ecole Polytechnique Federal Lausanne for providing the GaAs sample and Thierry Ondarcuhu from CEMES for AFM measurments.
\end{acknowledgments}


\providecommand{\noopsort}[1]{}\providecommand{\singleletter}[1]{#1}%

\subsection{Figure captions}
\begin{figure}[H]
\begin{center}
\includegraphics[width=180mm, trim=5 11 0 0,clip]{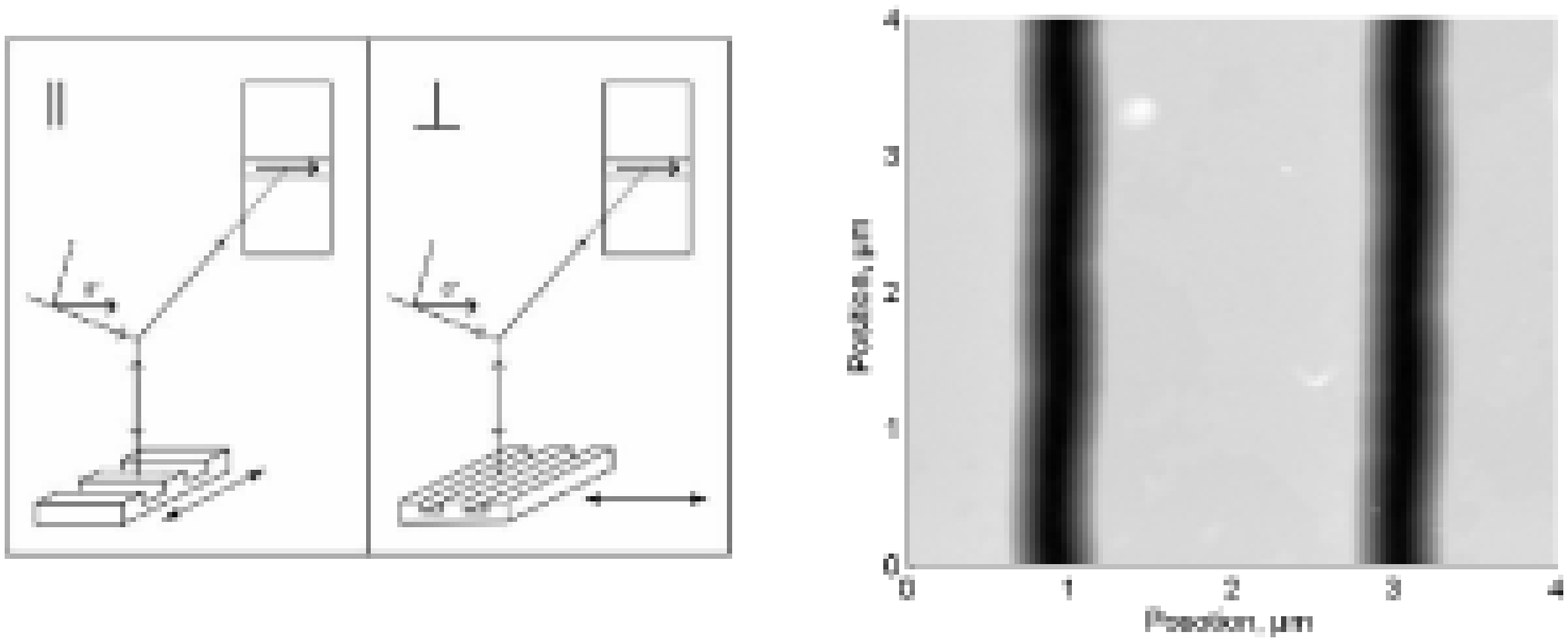}
\caption{}
\label{fig1}
\end{center}
\end{figure}

FIG. \ref{fig1}. Left side: schematic of two sample orientations with respect to spectrometer entrance slit and direction of polarization. The trenches of the substrate are either oriented parallel or perpendicular to the spectrometer entrance slit. Right side: AFM image of trenches (600nm wide, 500nm deep, distance between trenches: 2100nm).

\begin{figure}[H]
\begin{center}
\includegraphics[width=150mm, trim=5 11 0 0,clip]{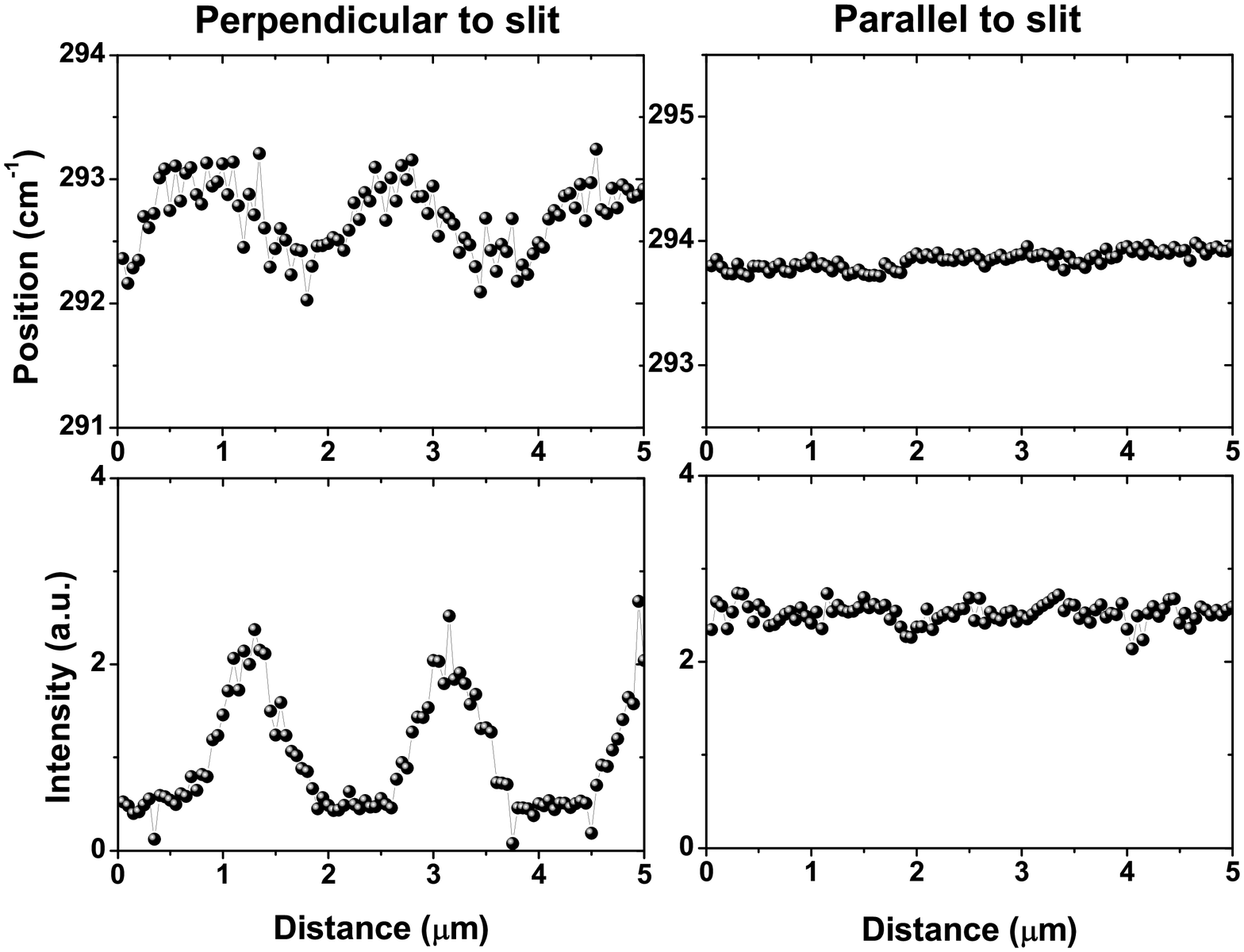}
\caption{}
\label{fig2}
\end{center}
\end{figure}

FIG. \ref{fig2}. Spectral position and intensity of Raman LO phonon in GaAs when scanning either perpendicular (left) or parallel (right) with respect to the spectrometer entrance slit.

\begin{figure}[H]
\begin{center}
\includegraphics[width=80mm, trim=5 11 0 0,clip]{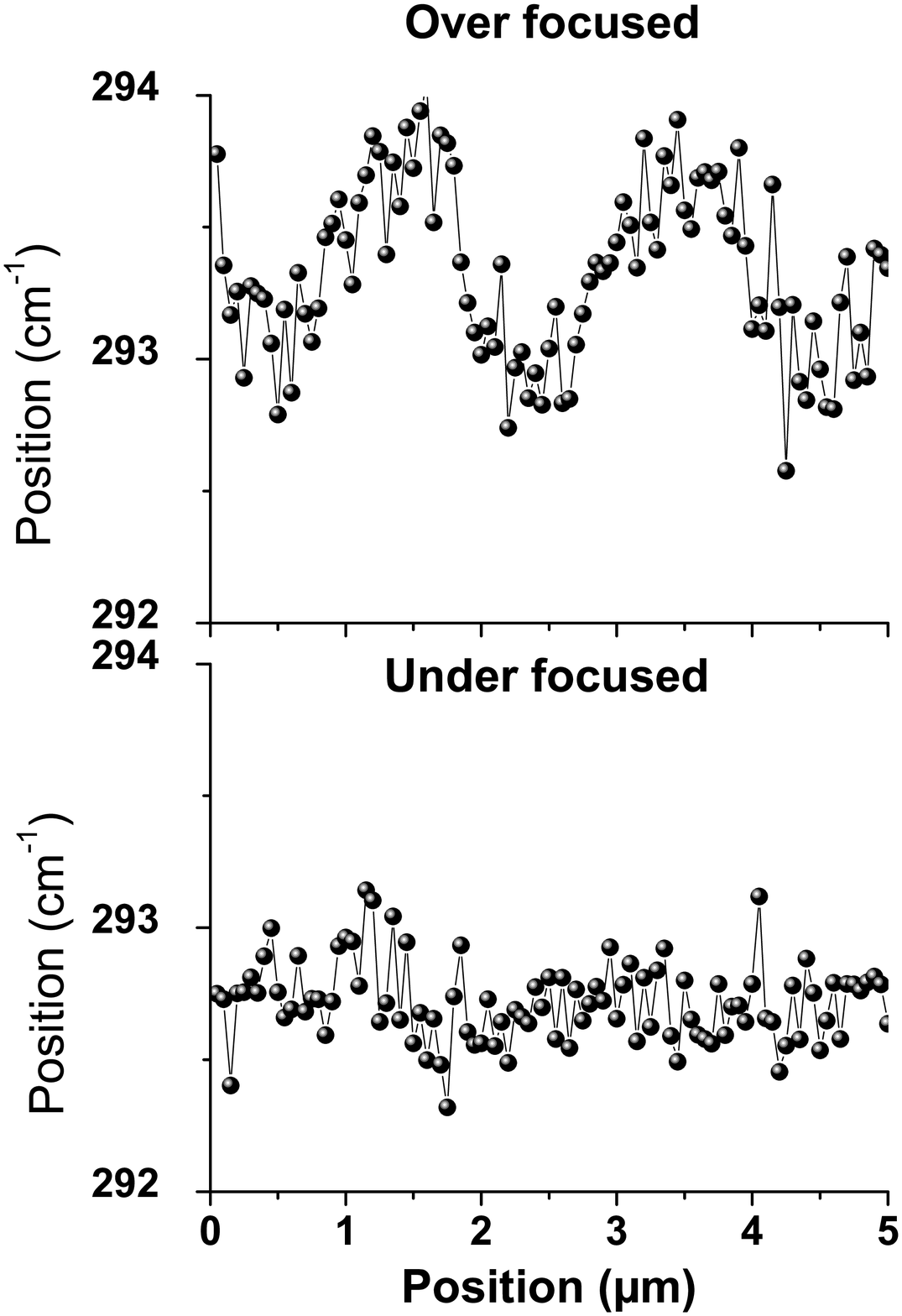}
\caption{}
\label{fig3}
\end{center}
\end{figure}

FIG. \ref{fig3}. Effect of over and under focusing when scanning perpendicular to the spectrometer entrance slit. The spot size after defocusing has been double the size when compared to the spot size when focused.

\end{document}